\documentclass[amsmath,superscriptaddress,reprint,aps,showkeys]{revtex4-2}
\usepackage{graphicx}
\usepackage{dcolumn}
\usepackage{bm}
\usepackage{hyperref}
\usepackage{siunitx}
\usepackage{color, soul}

\setstcolor{red}

\begin{document}
	\title{Ultrafast acoustic modulation of second-harmonic generation in monolayer transition metal dichalcogenides}

      \author{Takumi~Yamamoto}
    \affiliation{Department of Physics, Faculty of Science and Technology, Keio University, Yokohama, 223-8522, Japan}

    \author{Hidetoshi~Kanzawa}
    \affiliation{Department of Physics, Faculty of Science and Technology, Keio University, Yokohama, 223-8522, Japan}

    \author{Yuta~Takahashi}
    \affiliation{Department of Physics, Faculty of Science and Technology, Keio University, Yokohama, 223-8522, Japan}

    \author{Hajime~Kumazaki}
    \affiliation{Department of Physics, Faculty of Science and Technology, Keio University, Yokohama, 223-8522, Japan}

    \author{Jiang~Pu}
    \affiliation{Department of Physics, Institute of Science Tokyo, Tokyo, 152-8551, Japan}

    \author{Shinichi~Watanabe}
    \affiliation{Department of Physics, Faculty of Science and Technology, Keio University, Yokohama, 223-8522, Japan}

	\author{Shun~Fujii}
	\email[Corresponding author.]{shun.fujii@phys.keio.ac.jp}
 	\affiliation{Department of Physics, Faculty of Science and Technology, Keio University, Yokohama, 223-8522, Japan}
    
\begin{abstract}
High-speed modulation and deterministic control of optical nonlinear processes in nanomaterials are essential for realizing future nanoscale optoelectronic devices. Applying strain is a ubiquitous and versatile approach to deform atomically thin materials, allowing direct modification of their electronic and optical properties. Yet, strain engineering of nonlinear processes has so far relied predominantly on static approaches, which inherently limit modulation speed, reproducibility, and device scalability.
Here, we demonstrate ultrafast acoustic modulation of second-harmonic (SH) generation in monolayer transition metal dichalcogenides using surface acoustic waves (SAWs). By employing a fully phase-synchronized SH measurement combined with stroboscopic surface displacement detection, we directly visualize dynamic SH modulation at a frequency of 226 MHz. Moreover, theoretical modeling and determination of photoelastic coefficients enable quantitative extraction of the SAW-induced dynamic strain. Our results establish a direct link between acoustic fields and optical nonlinearities, providing a robust platform for dynamic strain engineering in two-dimensional nanophotonic devices. 


\it{This is the author's manuscript version of an article published in Nano Letters.  The version of record is available via DOI: \url{https://doi.org/10.1021/acs.nanolett.6c00268}.}
\end{abstract}

	\maketitle

van der Waals two-dimensional (2D) materials have emerged as promising platforms for nonlinear optics since the first discovery of graphene in 2004~\cite{Novoselov2004}. Despite their ultimately thin, atomically layered structures, 2D materials have demonstrated strong potential as nonlinear media owing to their giant nonlinear susceptibilities, which are comparable to those of commonly used bulk nonlinear crystals~\cite{Autere2018,You2019}. In particular, monolayer transition metal dichalcogenides (TMDCs) are at the forefront of 2D materials for nonlinear optical applications, as excitonic resonances combined with broken inversion symmetry substantially enhance second-order nonlinear optical processes~\cite{Kumar2013,Li2013,Malard2013}. Moreover, owing to their mechanical flexibility and robustness, the integration of monolayer TMDCs has emerged as a promising strategy to functionalize a wide range of photonic platforms, including optical microresonators~\cite{Fang2022,Fujii2024}, optical fibers~\cite{Zuo2020,Ngo2022}, and integrated waveguides~\cite{Chen2017,Mooshammer2024}.

Deterministic control of nonlinear optical processes in 2D materials is essential for emerging applications such as efficient nonlinear frequency converters~\cite{Wang2018,Klimmer2021}, quantum sensing~\cite{Fang2024}, and probing ultrafast coherent phenomena~\cite{Herrmann2025,Klimmer2026}. While efficient control of harmonic generation in 2D materials has been demonstrated using multilayer stacking~\cite{Hsu2014}, external electric fields~\cite{Seyler2015,Shree2021}, or all-optical modulation schemes~\cite{Wang2021,Klimmer2021}, mechanical strain provides an additional and highly effective control channel, as strain directly modifies the electronic band structure and optical properties of TMDCs~\cite{Peng2020}.

Indeed, polarization-resolved second-harmonic generation (SHG) measurements have been widely employed not only to identify the crystal orientation and layer number of honeycomb-lattice 2D materials~\cite{Kumar2013,Li2013}, but also to evaluate strain~\cite{Mennel2018,Kourmoulakis2024,Xing2024} and morphology~\cite{Zhao2016,Kaneda2025} in these crystals, owing to the strong sensitivity of the second-order nonlinear susceptibility to lattice deformation. These facts raise SHG as a powerful probe of strain in monolayer TMDCs. Despite this capability, strain-induced control of SHG has thus far been predominantly explored under static strain conditions. Such approaches may suffer from intrinsic limitations when implemented in device platforms, including slow response speeds, limited reproducibility, and poor repeatability under cyclic operation. These drawbacks significantly restrict their applicability for high-speed nonlinear and quantum optoelectronic devices~\cite{ReserbatPlantey2021}. 

To overcome these limitations, dynamic strain engineering using surface acoustic waves (SAWs) offers a promising strategy. Rayleigh SAWs, propagating along piezoelectric surfaces with coupled longitudinal and transverse motions, enable high-frequency and electrically controllable strain modulation. In monolayer TMDCs, SAWs have been utilized to modulate photoluminescence (PL)~\cite{Datta2021}, exciton transport~\cite{Datta2022}, and carrier dynamics~\cite{Nysten2024} through dynamic strain, demonstrating a powerful route for high-speed and reversible control of material properties.

In this work, we demonstrate microwave-frequency ultrafast acoustic modulation of SHG in monolayer TMDCs by employing surface acoustic waves (SAWs). By introducing periodic dynamic strain with electrically controlled phase and frequency, our approach overcomes the intrinsic limitations of static strain control and enables high-speed, repeatable, and quantitative modulation of nonlinear optical responses in 2D materials. Using a fully phase-synchronized polarization-resolved SH (PSH) measurement combined with stroboscopic surface displacement detection, we directly visualize the real-time modulation of SH light in response to dynamic strain at a frequency of 226~MHz with an intensity modulation depth as large as $\sim19$\%. Furthermore, we quantitatively extract the SAW-induced strain magnitude by modeling dynamic SHG modulation and by independently determining the photoelastic tensor elements through static strain experiments performed on four representative TMDC crystals. This study establishes a direct link between acoustic fields and SHG in 2D materials, opening a new avenue toward ultrafast and deterministic strain engineering of nonlinear optical processes in miniaturized device platforms.

\begin{figure*}
	\centering
		\includegraphics[width=0.9\textwidth]{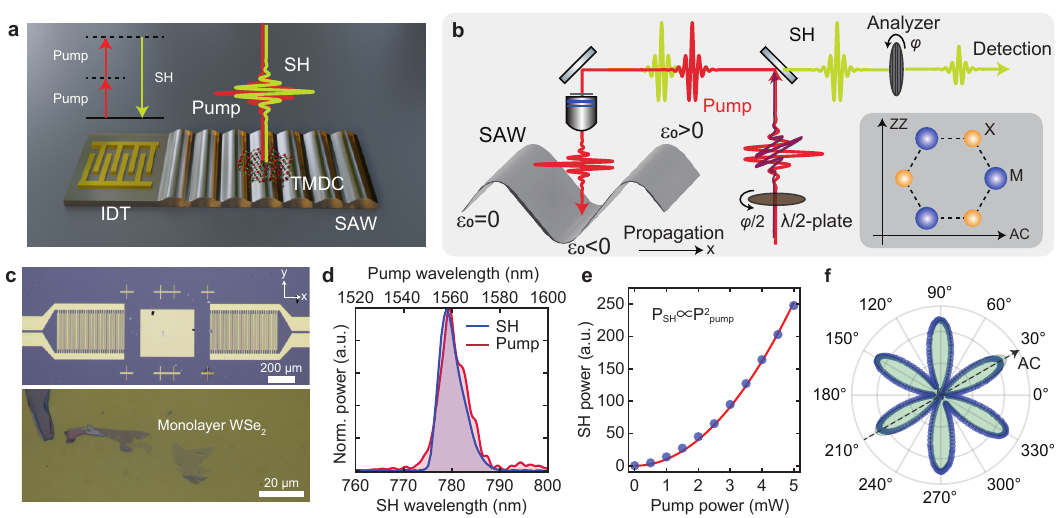}
		\caption{\label{Fig_concept} Concept of acoustic modulation of second-harmonic generation and basic characterization in monolayer TMDCs. (a) Conceptual illustration of a monolayer TMDC integrated with a SAW device. (b) Schematic of a polarization-resolved SH measurement under SAW excitation. The inset indicates the definition of the crystal orientation. AC: armchair; ZZ: zigzag. (c) Optical micrographs of a fabricated device with a transferred monolayer $\mathrm{WSe_2}$. (d) Optical spectra of the fundamental pump and the generated SH signal. (e) Pump power dependence of the SH intensity, showing a quadratic behavior characteristic of SHG. (f) Polarization-resolved SH intensity (flower pattern) used to determine the crystal orientation of the monolayer TMDC.}
\end{figure*}

Figure~\ref{Fig_concept}a presents a conceptual illustration of a monolayer TMDC–integrated SAW device, which serves as an acoustic modulation platform for second-harmonic (SH) light. Electrically driven Rayleigh SAWs generate dynamic strain fields with frequencies ranging from several hundreds of megahertz to a few gigahertz, which mechanically interact with the monolayer TMDC. A schematic of the SAW-induced SHG modulation setup is shown in Fig.~\ref{Fig_concept}b. In this configuration, PSH microscopy is employed to probe the influence of the acoustically induced strain on the SH intensity. For Rayleigh-type SAWs, the valley and apex of the acoustic wave correspond to compressive and tensile strain, respectively. Figure~\ref{Fig_concept}c shows optical micrographs of a fabricated device, in which a monolayer $\mathrm{WSe_2}$ is positioned between interdigital transducers (IDTs) that generate SAWs at a frequency of 226~MHz.

\begin{figure*}
	\centering
		\includegraphics[width=0.9\textwidth]{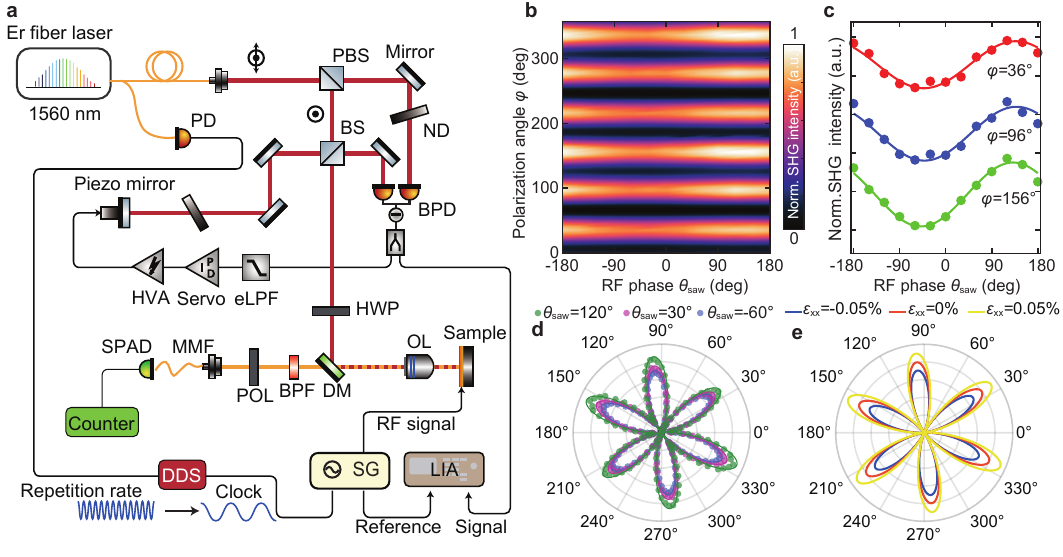}
		\caption{\label{Fig_setup_shg} Experimental setup and phase-synchronized SHG microscopy and acoustic modulation driven by surface acoustic waves. (a) Experimental setup. A 1560~nm femtosecond pulse laser is used as the excitation source, and polarization-resolved SH microscopy is performed under phase synchronization between the pulse repetition rate and the SAW frequency. A stabilized Michelson interferometer is implemented for surface displacement measurements. (B)PD, (balanced-)photodetector; PBS, polarizing beam splitter; BS, beam splitter; ND, neutral density filter; DM, dichroic mirror; OL, objective lens; BPF, band-pass filter; MMF, multimode fiber; SPAD, single-photon avalanche detector; Counter, time-tagger; DDS, direct digital synthesizer; SG, signal generator; LIA, lock-in amplifier; eLPF, electrical low-pass filter; HVA, high-voltage amplifier.   (b) Polarization-resolved mapping of the normalized SH intensity as a function of the RF phase in a monolayer $\mathrm{WSe_2}$.   (c) Acoustically modulated SH intensity for three representative AC directions.   (d) Polarization-resolved SH flower patterns at three different RF phases, corresponding to the maximum, median, and minimum SH intensities.   (e) Simulated flower patterns under different uniaxial strain amplitudes reproducing the experimental observations.}
\end{figure*}

We first perform a fundamental characterization of SHG from a monolayer $\mathrm{WSe_2}$. A femtosecond pulse laser is used as a fundamental pump source, and emitted SH light at a wavelength of $\sim780$~nm is collected and analyzed here. Optical spectra and the pump power dependence of the SH light are shown in Fig.~\ref{Fig_concept}d,e. The observed quadratic dependence on the pump power provides direct evidence of a typical SHG process. Under PSH measurements, a six-fold symmetric intensity pattern, often called a flower pattern, is observed. This enables the determination of crystal orientation, as the SH light is polarized along the armchair (AC) direction and is proportional to $\cos^2(3\varphi-3\delta)$, where $\varphi$ is the pump polarization angle and $\delta$ is the rotation of the AC direction relative to the $p$-polarization ($x$-axis) in the laboratory frame. As an example shown in Fig.~\ref{Fig_concept}f, the AC direction is determined to be $\delta = 29.9^\circ + n \times 60^\circ$, where $n$ is an integer. More importantly, the flower pattern provides valuable information on intrinsic and extrinsic strain through deviations from the ideal six-fold symmetry in TMDCs~\cite{Mennel2018,Xing2024,Liang2017}.

\begin{figure*}
	\centering
		\includegraphics[width=0.9\textwidth]{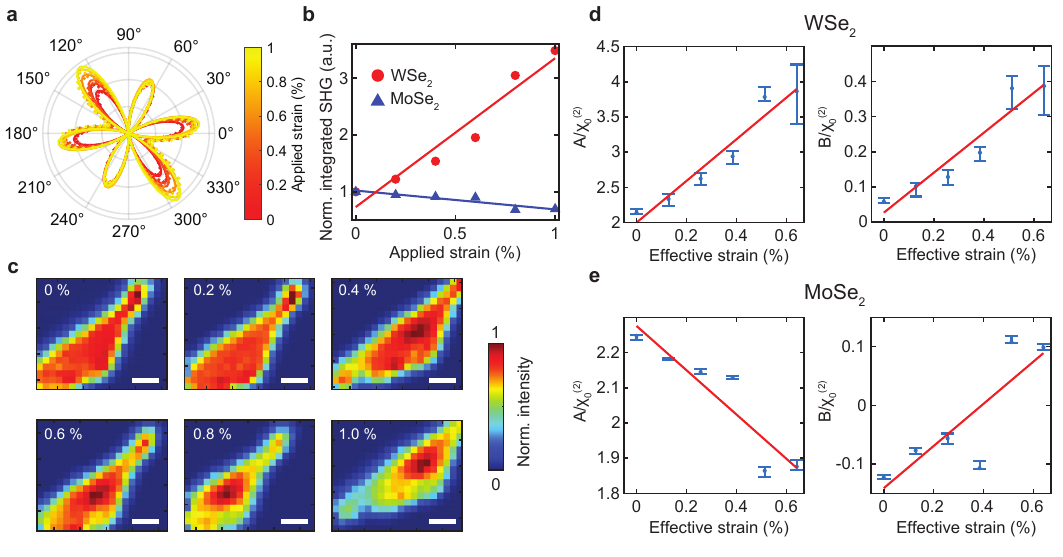}
		\caption{\label{Fig_static} Photoelastic coefficient estimation under static strain.  (a) Polarization-resolved SHG intensity normalized to the value at $\delta=66^\circ$ in monolayer $\mathrm{WSe_2}$. Tensile strain is applied along the horizontal direction. (b) Normalized integrated SH intensity as a function of the applied tensile strain, showing distinct strain-dependent trends for monolayer $\mathrm{WSe_2}$ and $\mathrm{MoSe_2}$.  (c) Spatial mapping of the SH intensity at different applied strain levels for monolayer $\mathrm{WSe_2}$. The scale bar is 2~\textmu m.  (d,e) Strain dependence of $A/\chi_0^{(2)}$ and $B/\chi_0^{(2)}$ for monolayer $\mathrm{WSe_2}$ and $\mathrm{MoSe_2}$, respectively, plotted using the effective (calibrated) strain.}
\end{figure*}

Visualizing the influence of ultrafast acoustic strain on the SH process is challenging because the spatial profile of the SAW is generally difficult to resolve due to its high-frequency oscillatory motion. We therefore utilize a fully phase-synchronized optical SH spectroscopy technique combined with surface displacement measurements based on a pulsed-laser Michelson interferometer~\cite{Shao2022,Maezawa2024}. A detailed experimental setup is presented in Fig.~\ref{Fig_setup_shg}a. By establishing a phase-locking condition between the SAW frequency $f_\mathrm{SAW}$ and an integer multiple of the pulse repetition rate $f_\mathrm{rep}$, the SH signals and the out-of-plane surface vibration are simultaneously monitored in a stroboscopic manner (see Methods for details). As a result, scanning the RF phase using a signal generator enables SAW phase-dependent PSH measurements, allowing us to visualize SH modulation induced by dynamic strain. It should be noted that the RF phase corresponds to the relative phase set by the signal generator and is preserved as long as the phase-locking condition is maintained.

Figure~\ref{Fig_setup_shg}b shows a PSH intensity map as a function of the RF phase $\theta_\mathrm{SAW}$, i.e., the SAW phase, for a monolayer $\mathrm{WSe_2}$ sample. The propagation direction of the SAW is along $x$-axis, and the RF driving power is 30~dBm. Dynamic modulation of the SH intensity is clearly observed over a SAW cycle, where the variations for all AC directions are well fitted by sinusoidal functions (Fig.~\ref{Fig_setup_shg}c). We also highlight the flower patterns at $\theta_\mathrm{SAW}=120^\circ$, $30^\circ$, and $-60^\circ$, which correspond to the maximum, median, and minimum SH intensity, respectively, as shown in Fig.~\ref{Fig_setup_shg}d. Surprisingly, the intensity modulation depth reaches up to $\sim19$\% for $\varphi=156^\circ$ despite the microstrain induced by SAWs. 

From these observations, nevertheless, we deduce that the local acoustic strain can modulate the SH intensity through a strain-dependent nonlinear susceptibility. The strain-dependent SHG intensity at the pump polarization direction $\varphi$ is 
defined as {\cite{Lyubchanskii2000,Mennel2018a,Mennel2018},
\begin{equation}
I_{\parallel}^{(2)}(2\omega) \propto \left( A \cos(3\varphi - 3\delta) + B \cos(2\theta + \varphi - 3\delta) \right)^2, \label{eq:shg_strain}
\end{equation}
where $A = (1-\nu)(p_1 + p_2)(\varepsilon_{xx}+\varepsilon_{yy}) + 2\chi_0^{(2)}$ and $B = (1+\nu)(p_1 - p_2)(\varepsilon_{xx}-\varepsilon_{yy})$, and $\theta$ is the principal strain angle. The factors $A$ and $B$ include the Poisson's ratio $\nu$, where $p_1$ and $p_2$ are non-zero photoelastic elements, and $\varepsilon_{xx}$ and $\varepsilon_{yy}$ denote the applied strain components. Physically, the photoelastic tensor quantifies how lattice deformation modifies the nonlinear optical response. The first term with $A$ changes the overall SHG intensity, whereas the second term with $B$ causes the symmetry breaking of the six-fold pattern~\cite{Mennel2018}. We also confirm that Eq.~\ref{eq:shg_strain} qualitatively reproduces the strain-modulated flower patterns as shown in Fig.~\ref{Fig_setup_shg}e. Additional simulations for different strain combinations are provided in Supporting Information, Section~1.

Since a particular interest here is ultrafast dynamic SAW strain, we modify the factors $A$ and $B$ as
\begin{gather}
A = (1-\nu)(p_1 + p_2)(\varepsilon_{0} \sin\theta_\mathrm{SAW})+ 2\chi_0^{(2)}, \label{eq:shg_strain2} \\
B = (1+\nu)(p_1 - p_2)(\varepsilon_{0} \sin\theta_\mathrm{SAW}), \label{eq:shg_strain3}
\end{gather}
where $\varepsilon_{0}$ is the maximum dynamic strain induced by the SAW. It should be noted that $\varepsilon_{0}$ represents the uniaxial in-plane strain component along the SAW propagation direction, $\varepsilon_{xx}$, which provides the leading contribution to the Rayleigh-type SAW-induced strain field~\cite{wei2022theory}. These relations indicate that an accurate determination of the photoelastic coefficients $p_1$ and $p_2$ is essential for analyzing dynamic strain. Once $p_1$ and $p_2$ are reliably determined, Eq.~\ref{eq:shg_strain} can be used to fit the SAW phase-dependent PSH intensity shown in Fig.~\ref{Fig_setup_shg}b, thereby enabling direct verification of SAW-induced dynamic SHG modulation and quantitative estimation of the strain magnitude. This motivates the following experiments, in which we independently estimate $p_1$ and $p_2$ to establish a quantitative link between the measured SH response and the underlying SAW-driven strain.

\begin{figure*}
	\centering
		\includegraphics[width=0.9\textwidth]{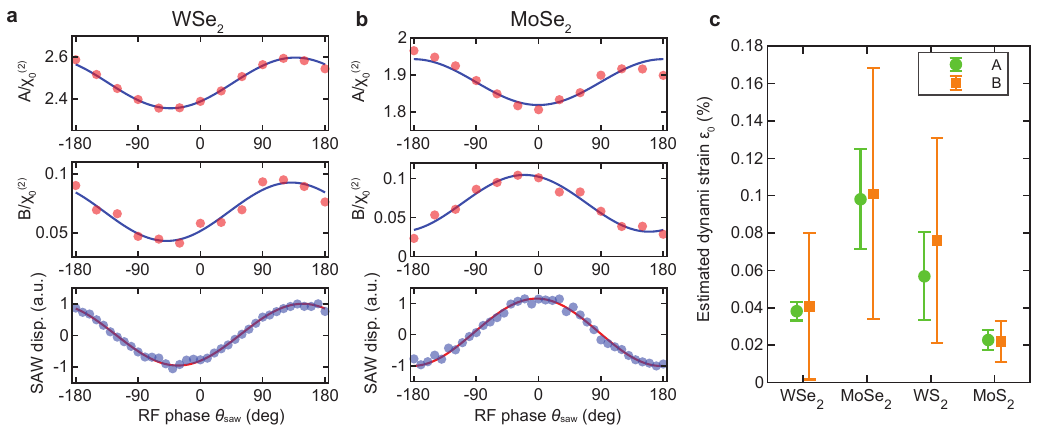}
		\caption{\label{Fig_dynamic} Phase-dependent SHG response and quantitative estimation of SAW-induced dynamic strain. (a,b) Extracted $A/\chi_0^{(2)}$ and $B/\chi_0^{(2)}$, plotted as functions of the RF phase together with the measured SAW surface displacement, for monolayer $\mathrm{WSe_2}$ and $\mathrm{MoSe_2}$. The solid lines represent sinusoidal fits to the experimental data (circles). The valley and apex of the SAW displacement correspond to maximized compressive and tensile strain, respectively. (c) Estimated dynamic strain extracted independently from the $A/\chi_0^{(2)}$ and $B/\chi_0^{(2)}$ coefficients for four representative TMDC monolayers.}
\end{figure*}

We then conduct separate PSH measurements for monolayer TMDC samples transferred onto flexible substrates (see Methods for details). Tensile strain ($\varepsilon_{xx} > 0$) is applied to the monolayers along the horizontal ($0^\circ$) axis using custom-designed bending jigs, where the applied strain is determined by the relation $\varepsilon_{xx} = d/2R$, with $d$ denoting the substrate thickness and $R$ the curvature radius~\cite{Pu2021} (Supplementary Fig.~1). Using this approach, we systematically apply stepwise tensile strain to monolayer samples and monitor the resulting evolution of the normalized SH flower patterns. A representative result for $\mathrm{WSe_2}$ is shown in Fig.~\ref{Fig_static}a, where clear asymmetric modification of the flower pattern is observed with increasing strain, reflecting strain-induced changes in the SHG response. Figure~\ref{Fig_static}b also presents the integrated intensity evolution under tensile strain, where the SHG signal is integrated over all polarization angles. We observe an opposite trend for $\mathrm{WSe_2}$ and $\mathrm{MoSe_2}$, which suggests that the intensity responses are significantly different among different TMDCs.

By fitting the measured flower patterns at each strain level using Eq.~\ref{eq:shg_strain}, we extract the strain dependence of the coefficients $A$ and $B$, from which the corresponding slopes with respect to the applied strain are obtained. To ensure statistically reliable determination of the photoelastic tensor elements, we perform spatial mapping of the PSH response as shown in Fig.~\ref{Fig_static}c, where the normalized intensity is presented only. The nonuniform intensity evolution at higher strain levels can be attributed to the large curvature radius of the bending jigs. The fitting procedure is applied to the same spatial regions identified from the mapping measurements, allowing us to extract consistent strain-dependent slopes of $A$ and $B$ with minimized influence of spatial inhomogeneity. This approach enables robust and quantitative estimation of the photoelastic coefficients for monolayer TMDCs. 

It is worth mentioning that slippage should be considered when estimating actual strain levels in bending experiments. We therefore conduct PL spectrum measurements using the same jig to evaluate effective strain levels (Supporting Information, Section~2). Taking the slipping effects and the strain efficiency of $\sim$64\% during the bending procedures into account, we obtain 
${\partial (A/\chi_0^{(2)})}/{\partial \varepsilon_{xx}} = 2.96\pm0.51$ and 
${\partial (B/\chi_0^{(2)})}/{\partial \varepsilon_{xx}} = 0.568\pm0.13$ for monolayer $\mathrm{WSe_2}$ as shown in Fig.~\ref{Fig_static}d.  For monolayer $\mathrm{MoSe_2}$, the corresponding values are 
${\partial (A/\chi_0^{(2)})}/{\partial \varepsilon_{xx}}  = -0.630\pm0.18$ and  ${\partial (B/\chi_0^{(2)})}/{\partial \varepsilon_{xx}}  = 0.360\pm1.17$ (Fig.~\ref{Fig_static}e). The slight deviation from the linear trend may arise from local strain redistribution or partial interfacial slip between the TMDC and the substrate. From these slopes, the photoelastic combinations $(p_1+p_2)/\chi_0^{(2)}$ and $(p_1-p_2)/\chi_0^{(2)}$ are evaluated to be $3.66\pm0.46$ and $0.478\pm0.46$ for $\mathrm{WSe_2}$, and $-0.818\pm0.19$ and $0.292\pm0.19$ for $\mathrm{MoSe_2}$, respectively. We find that the sign of the $A$-term slope, corresponding to that of $(p_1+p_2)/\chi_0^{(2)}$, is opposite between $\mathrm{WSe_2}$ and $\mathrm{MoSe_2}$. This sign reversal is consistent with the different trend in SH intensity under tensile strain (Fig.~\ref{Fig_static}b), indicating a unique strain response of SHG processes in TMDCs, which may be influenced by material-specific excitonic resonances and their strain-induced spectral shifts relative to the excitation wavelength~\cite{Mennel2018,Guan2025}.

We finally extract the SAW phase dependence of the two coefficients $A$ and $B$ by fitting the polarization-resolved SH flower patterns at each SAW phase in the PSH map (Fig.~\ref{Fig_setup_shg}b for $\mathrm{WSe_2}$) using Eqs.~\ref{eq:shg_strain2}-\ref{eq:shg_strain3}. The results of the fittings are shown separately for monolayer $\mathrm{WSe_2}$ and $\mathrm{MoSe_2}$ in Fig.~\ref{Fig_dynamic}a,b, and the measured SAW surface displacements are also plotted together for direct comparison. Both $A/\chi^{(2)}_0$ and $B/\chi^{(2)}_0$ exhibit clear sinusoidal modulation synchronized with the SAW displacement, indicating that the observed SHG modulation originates from dynamic acoustic strain. Using the same measurement and fitting procedure, we further apply this analysis to two other representative TMDCs, $\mathrm{WS_2}$ and $\mathrm{MoS_2}$. The corresponding results are provided in Supporting Information, Section~3, and the values are summarized in Supplementary Table~1.

Importantly, the independent fitting of $A/\chi^{(2)}_0$ and $B/\chi^{(2)}_0$ using Eq.~\ref{eq:shg_strain2}-\ref{eq:shg_strain3} yields consistent estimates of the SAW-induced strain magnitude, with the strain values extracted from $A$ and $B$ falling within almost the same quantitative range. We note that the influence of the pre-existing initial strain is explicitly accounted for in our analysis and can be separated from the SAW-induced modulation. A detailed analysis and error propagation of the measurement are described in Supporting Information, Section~4-5. Figure~\ref{Fig_dynamic}c summarizes the maximum dynamic strain $\varepsilon_0$ estimated from the corresponding $A$ and $B$ fittings for four TMDC materials. It is noteworthy that the estimated strain values of up to $\sim0.02-0.1$\% are comparable to strain magnitudes reported for similar SAW devices using $\mathrm{LiNbO_3}$ substrates~\cite{Lazic2019,Datta2022}.

Not only the consistency among different TMDC materials, but also the quantitative agreement with previously reported SAW-induced strain values confirms the general applicability of our approach across other 2D materials and establishes a reliable framework for high-precision evaluation of SAW-induced dynamic strain. We note that the relatively large uncertainty in the estimated strain primarily originates from the uncertainties in the photoelastic coefficients $p_1$ and $p_2$. Therefore, further refinement in the determination of $p_1$ and $p_2$ is expected to substantially improve the accuracy of the extracted dynamic strain.

In conclusion, we have demonstrated acoustic modulation of SHG in monolayer TMDCs driven by dynamic strain. Fully phase-synchronized PSH microscopy enables clear visualization of dynamic SH intensity modulation at a frequency of 226~MHz, originating from the high strain sensitivity of the second-order optical nonlinearity in monolayer TMDCs. Stroboscopic surface displacement measurements provide direct evidence of the in situ response of SH processes to dynamically applied strain. Furthermore, accurate determination of the photoelastic tensor through static strain experiments allows quantitative verification of the dynamic strain magnitude induced by the SAW.

Our results envision ultrafast modulation of SHG in monolayer TMDCs, representing the first demonstration that dynamic acoustic strain, rather than static strain, offers an effective and versatile route for arbitrary control of nonlinear optical responses in 2D materials. While static strain enables larger strain amplitudes, SAW-based modulation provides electrically controlled, high-frequency periodic strain with superior temporal precision and device compatibility. Owing to the scalability of SAW frequencies into the gigahertz regime~\cite{Li2015}, the present approach can, in principle, be extended toward higher-frequency modulation, with an intrinsic modulation timescale defined by the SAW period (sub-nanosecond in the GHz regime). However, the practical upper frequency limit of this method is ultimately constrained by the ratio between the SAW wavelength and the laser spot size, since significant spatial phase variation within the excitation area may lead to partial signal averaging. Such dynamic control of second-order optical nonlinearity is expected to play an important role not only in classical nonlinear optics, but also in quantum light generation, such as single photon emission~\cite{Lazic2019,Parto2021} and entangled photon generation~\cite{Weissflog2024}, based on rich nonlinear processes in 2D materials, where high-speed and deterministic control is essential.

Furthermore, the high strain sensitivity of SHG demonstrated here establishes nonlinear optical responses of TMDCs as a powerful and quantitative probe for detecting microscopic SAW-induced dynamic strain. This functionality, serving both as an ultrafast nonlinear optical modulator and as a sensitive nanoscale strain sensor, underscores the broad applicability of our approach to strain-engineered photonics and acousto-optic technologies based on 2D van der Waals materials.

\section*{Methods}

{\it Device fabrication}\\
The SAW devices were fabricated on 128$^\circ$ Y $\mathrm{LiNbO_3}$ substrates using a standard photolithography process. The IDTs with a pitch of  $\Lambda=8.42$~\textmu m ($2\Lambda=\lambda_\mathrm{SAW}$) were patterned using laser lithography (DWL66+, Heidelberg Instruments). The sample was then exposed to electron-beam (EB) evaporation, and a 70-nm Au layer was stacked on a 3-nm Ti layer. As a second lithography process, a 400~\textmu m by 350~\textmu m square pad was patterned between two IDTs, and then 3-nm-Ti/70-nm-Au layers were evaporated, followed by deposition of a 90-nm silicon dioxide layer. The lift-off processes were performed by immersing the samples in a polar solvent, N-methylpyrrolidone (NMP). The Au pad was used to screen out strong SHG processes from a $\mathrm{LiNbO_3}$ substrate, and the $\mathrm{SiO_2}$ layer enhanced the collection efficiency of SHG signals.

Monolayer TMDC flakes were mechanically exfoliated from bulk crystals (HQ Graphene) using cellophane tape and transferred from a PDMS stamp (PF-40x40-0170-X4, Gel-Pak) to the square pad on a SAW device using a dry-transfer technique. The IDTs were electrically connected to external sources using wire bonding. The SAW transmission was characterized by an $S_{21}$ measurement using a vector network analyzer (Supplementary Fig.~2).

{\it Characterization of second-harmonic generation}\\
A home-built mode-locked femtosecond fiber laser at a center wavelength of 1560~nm was used as a pump source, where the repetition rate and the pulse width were 75.4~MHz and $\sim$250 fs, respectively. The pump beam was focused to a spot size of 2~\textmu m on the sample plane using a 50X objective lens (LCPLN50XIR, Olympus). The average pump power was approximately 3~mW, corresponding to a pulse energy density of 2.5~$\mathrm{mJ/cm^2}$.  A half-wave plate (WPH05M-1550, Thorlabs) was utilized to control the pump polarization angle $\varphi$. For polarization-resolved measurements, a linear polarizer (\#70998, Edmund Optics) was inserted before fiber coupling, where the detection polarization was always set parallel to the pump polarization. Both the waveplate and the polarizer were mounted on motorized rotator stages, allowing rapid and precise control of the angles. After spectral filtering using a band-pass filter (FBH780-10, Thorlabs), a fiber-coupled single-photon avalanche diode module (MPD) and a time-tagger (ID801, IDQ) were used to count SH light. 

{\it Probing surface displacement vibrations by pulsed laser interferometry}\\
Stroboscopic sampling using pulsed-laser interferometry was employed to measure the surface displacement induced by the SAW~\cite{Shao2022,Maezawa2024}. The laser pulse train probed the local SAW-induced displacement through an objective lens and was subsequently interferometrically combined with a reference arm in a Michelson interferometer. The length of the reference arm was coarsely adjusted using a manual translation stage, while a piezoelectric mirror was used to maintain equal optical path lengths. The interference signal detected by a balanced photodetector (PDB210C/M, Thorlabs) was split into two electrical paths: one was sent to a lock-in amplifier (SR830, SRS), and the other was used for closed-loop PID control to stabilize the interferometer at the quadrature point, thereby enhancing detection sensitivity and suppressing environmental path length fluctuations. A portion of the pump laser light was used to balance the optical powers incident on the detector. A mathematical description of the lock-in detection is provided in Supporting Information, Section~6.

As the SAW frequency $f_{\mathrm{SAW}}$ is set to an integer multiple of the pulse repetition rate ($f_{\mathrm{SAW}} = 3 \times f_{\mathrm{rep}}$), satisfying the phase-synchronization condition, the in-phase component of the lock-in output ($X$) directly reflects the out-of-plane surface displacement. In addition, a 10~MHz reference clock was generated from the photodetected pulse repetition rate using a direct-digital synthesizer, which disciplined the RF signal generator used for SAW excitation and ensured relative phase stability. For lock-in detection, a low-frequency amplitude modulation at 10~kHz was applied to the RF signal generator. While the in-phase lock-in signal cannot intrinsically distinguish between the peak and trough positions of the SAW due to a $\pi$-phase ambiguity, this ambiguity was uniquely resolved from the corresponding increase or decrease in SHG intensity under tensile strain.

{\it Estimation of photoelastic tensor by static strain experiments}\\
Uniaxial static strain was applied to  monolayer TMDCs by mechanically bending a 250-\textmu m-thick polyethylene terephthalate (PET) film (Lumirror, TORAY). Monolayers were mechanically exfoliated and transferred via a dry-transfer method to a cleaned film. For polarization-resolved spatial mapping, a sample stage was scanned across the measurement area.





\section*{Acknowledgments}
Parts of this study are supported by JSPS KAKENHI (JP24H01202, JP23H00262, JP25KJ2060, JP25H02153, JP24K21743, JP23K26165, JP21H05232, JP21H05236); MEXT Quantum Leap Flagship Program (Q-LEAP) (JPMXS0118067246); JST CREST (JPMJCR19J4); JST FOREST (JPMJFR223Z); the Precise Measurement Technology Promotion Foundation (PMTP-F); Iketani Science and Technology Foundation; the Hattori Hokokai Foundation; Inamori Foundation; the Fujikura Foundation; Keio University
Program for the Advancement of Next Generation Research Projects. This work was partly conducted at the AIST Nano-Processing Facility supported by the Nanotechnology Platform Program of the MEXT (JPMXP09-F-21-AT-0085). The authors thank Kai Yamaguchi and Kazuki Maezawa at Keio University for technical assistance.

\section*{Author contributions}
S.F. conceived the experiments and supervised the project. T.Y. and H.Ka. prepared samples and performed experiments. T.Y., H.Ka, and S.F analyzed data. Y.T. and H.Ku. assisted the development of experimental setups. J.P. helped static strain measurements. S.W. helped sample fabrication.
The manuscript was written by T.Y. and S.F. with comments and inputs from all authors.

 	\bibliographystyle{achemso}					
	\bibliography{shg_saw}


\end{document}